\documentclass{ametsoc_edit}
\bibpunct{(}{)}{;}{a}{}{,}

\title{\textbf{\large{Long-range persistence in global surface temperatures explained by linear multi-box energy balance models}}}

\authors{Hege-Beate Fredriksen\correspondingauthor{Hege-Beate Fredriksen, Department of Mathematics and Statistics, University of Troms{\o}, the Arctic University of Norway, N-9037 Troms{\o}, Norway}
and Martin Rypdal}

\affiliation{Department of Mathematics and Statistics, University of Troms{\o}, the Arctic University of Norway, Norway}
\email{hege-beate.fredriksen@uit.no}

\abstract{The temporal fluctuations in global mean surface temperature is an example of a geophysical quantity which can be described using the notions of long-range persistence and scale invariance/scaling, but this description has suffered from lack of a generally accepted physical explanation. Processes with these statistical signatures can arise from non-linear effects, for instance through cascade-like energy transfer in turbulent fluids, but they can also be produced by linear models with scale-invariant impulse-response functions. This paper demonstrates that on time scales from months to centuries, the scale-invariant impulse-response function of global surface temperature can be explained from simple linear multi-box energy balance models. This explanation describes both the scale invariance of the internal variability and the lack of a characteristic time scale of the response to external forcings. With parameters estimated from observational data, the climate response is approximately scaling in these models, even if the response function is not chosen to be scaling {\em a priori}. It is also demonstrated that the differences in scaling exponents for temperatures over land and for sea-surface temperatures can be reproduced by a version of the multi-box energy balance model with two distinct surface boxes.}

\begin{document}
\maketitle

\section{Introduction} \label{sec:intro}
Instrumental measurements and proxy reconstructions of Earth's surface temperatures show temporal variability on a range of different time scales \citep{Lovejoy:2015, Huybers:2006}. For the global mean surface temperature (GMST) the variability can be parsimoniously described as scale invariant, since the estimated power spectral densities (PSDs) are well approximated by power-laws $S(f) \sim 1/f^\beta$ from monthly to centennial scales \citep{Rypdal:2013}. The typical scaling exponent is $\beta \approx 1$, and the signals are well described as a so-called $1/f$-noise, or pink noise. Some of the low-frequency variability in the temperature records can be accounted for by the variability in the radiative forcing of the planet, but even the residual fluctuations are well described as a scaling stochastic process,  with a slightly lower exponent $\beta$. This suggests that scale-invariant dynamics is an intrinsic property of the climate system, a claim that is supported by the observation of scaling PSDs in unforced control runs of general circulation models (GCMs), on time scales from months to centuries \citep{Fredriksen:2016, Rybski:2008, Fraedrich:2003}. 

A signal with power-law PSD can be modelled as a stochastic process with long-range dependence (LRD), and examples of such processes are the fractional Gaussian noises (fGns) and the fractional autoregressive integrated moving average (FARIMA) models. Stochastic processes that exhibit LRD provide more accurate descriptions of the unforced GMST variability compared to the traditional red-noise models, such as the Ornstein-Uhlenbeck (OU) processes and the autoregressive processes of order 1 (AR(1)) \citep{Rypdalx2:2014}. The latter  are characterised by a single time scale, and are incapable of describing the multi-scale nature of the climate fluctuations. Despite this, the LRD processes are largely ignored by many climate scientists, and some consider LRD  to be an exotic and  redundant notion in climate science \citep{Mann:2011}. 

One of the aims of this paper is therefore to de-mystify the notion of LRD in the climate system by demonstrating that the observed phenomena can be produced by simple multi-box energy balance models (EBMs). With this, we demonstrate that the exotic physics may be no more than vertical heat conduction in the ocean, and that it is reasonable to think of LRD as an approximation to the linear  response of EBMs with multiple characteristic time scales. Only a few boxes are needed to obtain power-law PSDs on scales from months to centuries. We also demonstrate how we can construct box models that are consistent with the observation that the exponent $\beta$ is lower for land temperatures than for sea surface temperatures (SSTs) \citep{Fredriksen:2016}.

Only a few of the studies that analyze LRD in surface temperatures, focus on the mechanisms behind the  phenomenon (e.g. \cite{Fraedrich:2003}, \cite{Fraedrich:2004}, \cite{Blender:2006}, \cite{Franzke:2015}). Most treat LRD-processes merely  as  statistical models that fit well with data  \citep{Vyushin:2012, Rybski:2006, Franzke:2010}. Statistical inference for LRD processes requires special care to avoid the fallacy of circular reasoning, that is, falsely attributing trends in the forced signal to natural variability \citep{Benestad:2016}. Incautious trend-significance testing using LRD null models \citep{CohnandLins} have led some climate scientists to view LRD processes as exotic mathematical objects that somehow fit with the ``climate denier agenda'' \citep{Mann:2011, Benestad:2016}. This is paradoxical, since climate response models that exhibit LRD actually display more ``heating in the pipeline'', and compared with other response models they predict that emissions of greenhouse gases must be reduced earlier and more drastically to avoid dangerous anthropogenic influence \citep{Rypdal:2016mm, Rypdalx2:2014}. 

Other climate scientists consider scaling to be an inherent property of atmospheric turbulent flows and certain types of regime switching dynamics \citep{Lovejoy:2013, Franzke:2015}, and as such a signature of the nonlinearity of the underlying dynamics. In fact, \cite{Huybers:2006} hypothesize that the persistent scaling of surface temperatures observed on decadal to multicentennial scales is due to a nonlinear cascade driven by the seasonal forcing. They present a bicoherence spectrum in favor of this hypothesis, but the phase correlations that give rise to high bicoherence do not imply an effective nonlinear energy transfer between the seasonal and the multidecadal scales. We have also had problems in reproducing the bicoherence spectra reported in this paper. In a forthcoming paper we will examine this hypothesis in depth.

Since the ocean has a large heat capacity  compared to the atmosphere, the observation that ocean temperatures are more persistent than land temperatures \citep{Fraedrich:2003, Fredriksen:2016}, is an indication that the observed persistence in global temperature to a larger extent must be attributed to ocean heat content and ocean dynamics, and to a lesser extent to nonlinear processes in the atmosphere. This hypothesis is further strengthened by the results of \cite{Fraedrich:2003}, who find that only models with full ocean circulation show persistence on scales longer than about a decade. In the present  paper we model the slowly responding components of the climate system by including ``boxes'' that exchange heat with the more rapidly responding mixed layer. This is clearly an oversimplification of the ocean dynamics, but reproduces the multi-scale characteristics of the surface temperature response.

The paper is structured as follows. Sec.~\ref{sec:2}  discusses the construction of multi-box EBMs and their corresponding response functions, and in Sec.~\ref{sec:3} we demonstrate how the superposition of different response times can be used to approximate an LRD response. Furthermore, we estimate parameters and explore how the response of sea surface temperatures differ from the response of land temperatures. Sec.~\ref{sec:4} presents some concluding remarks.

\section{Multibox EBMs} \label{sec:2}
The simplest climate model we can imagine is the so-called one-box EBM for the global temperature,
\begin{equation}
C \frac{d\Delta T}{dt} = -\frac{1}{S_\text{eq}} \Delta T + \Delta F(t). \label{eq:1boxebm}
\end{equation}
In this equation, $C$ denotes the average heat capacity per square meter of the surface, $\Delta T$ the temperature anomaly relative to an equilibrium state, $S_\text{eq}$ the equilibrium climate sensitivity, and $\Delta F(t)$ is the forcing, i.e., the perturbation of effective radiative forcing from the initial equilibrium state $\Delta T=0$. As a response to a constant perturbation $\Delta F$ the temperature will reach a new equilibrium $\Delta T$, and the change in equilibrium temperature relative to the change in radiative forcing is equal to the equilibrium climate sensitivity, i.e.,
\begin{equation*}
S_\text{eq} = \frac{\Delta T}{\Delta F}. 
\end{equation*}
For a time-dependent forcing $\Delta F(t)$, the temperature $\Delta T(t)$ is given by a convolution integral
\begin{equation}
\Delta T(t) = \int_{-\infty}^{t} R(t-s) \Delta F(s) ds, \label{eq:convintegral}
\end{equation}
where the impulse response function is an exponentially decaying function with a characteristic time scale $\tau = C S_\text{eq}$:
\begin{equation}
R(t) = \frac{1}{C} e^{-t/\tau}.
\end{equation}

In the one-box model there is no heat exchange with the deep ocean, but this can be included by extending the model to also include a box with a larger heat capacity $C_2$. If the energy exchange between the upper and lower box is proportional to the temperature difference between the two boxes, we obtain what is known as the two-box EBM \citep{Geoffroy:2013one, Held:2010, Rypdal:2012, Caldeira:2013}:
\begin{align}
C_1 \frac{d\Delta T_1}{dt} &= -\frac{1}{S_{\text{eq}}} \Delta T_1 +\kappa_2 (\Delta T_2 - \Delta T_1) + \Delta F(t) \\
C_2 \frac{d\Delta T_2}{dt} &= - \kappa_2 (\Delta T_2 - \Delta T_1). 
\end{align}
The equations can be written on matrix-form: 
\begin{equation} 
 \mathbf{C} \frac{d\Delta \mathbf{T}}{dt} = \mathbf{K} \; \Delta \mathbf{T} + \Delta \mathbf{F(t)} \label{eq:compactNboxeq}
\end{equation}
where we introduce the notation  
\begin{equation*}
\mathbf{C} = 
 \begin{pmatrix}
  C_1 & 0 \\
  0 & C_2 
 \end{pmatrix}, \text{~}
\Delta \mathbf{T} =  \begin{pmatrix}
\Delta T_1 \\
\Delta T_2
 \end{pmatrix}, \text{~}
 \Delta \mathbf{F(t)} =\begin{pmatrix}
\Delta F(t) \\
0
\end{pmatrix}
 \end{equation*}
 and 
\begin{equation} \label{K}
 \mathbf{K} =  \begin{pmatrix}
 -(\kappa_1 + \kappa_2) & \kappa_2 \\
  \kappa_2 & -\kappa_2 
 \end{pmatrix}. 
\end{equation}
For convenience we denote $\kappa_1 = 1/S_\text{eq}$, but it should be noted that the physical meaning of $\kappa_1$ is different from $\kappa_2$. While $\kappa_2$ is a coefficient of heat transfer between two ocean layers, $\kappa_1$ is determined by the response of the outgoing long-wave radiation (OLR) to changes in the surface temperature. It includes all the atmospheric feedbacks and is sometimes referred to as the {\em equilibrium global climate feedback}  \citep{Armour:2013}, or simply the {\em feedback parameter}.

The natural generalization of the two-box model is to consider $N$ vertically distributed boxes. The model is formulated as in Eq.~(\ref{eq:compactNboxeq}), with $\Delta \mathbf{T}$ and $\Delta \mathbf{F}$ being $N$-vectors, and $\mathbf{C}$ and $\mathbf{K}$ being $N\times N$ matrices. $\mathbf{C}$ will be a diagonal matrix with the heat capacities of each box along the diagonal, and $\mathbf{K}$ will be a tridiagonal matrix. We note that when $N$ is large, this model set-up can approximate a vertical diffusion model. 

In an $N$-box EBM the temperature $N$-vector can be written using matrix-exponential notation:
\begin{equation*}
\Delta {\bf T}(t) = \int_{-\infty}^t e^{(t-s){\bf A}} \Delta {\bf F}(s)\,ds,  \text{ with } {\bf A}={\bf C}^{-1} {\bf K}, 
\end{equation*}
and it follows that the 
surface temperature is given by a convolution integral similar to the one in Eq.~(\ref{eq:convintegral}), but where the impulse response function is now a weighted sum of $N$ exponentially decaying functions:
\begin{equation}
R(t)=\left( e^{t{\bf A}} \right)_{11} = \sum_{k=1}^N b_k e^{-t/\tau_k}.  \label{eq:responsef}
\end{equation}
The characteristic time scales are defined as $\tau_k = -1/\lambda_k, \; k=1,...,N$, where $\lambda_k$ are the eigenvalues of the matrix $\mathbf{C}^{-1}\mathbf{K}$. Since $-\mathbf{K}$ is symmetric and positive definite, the eigenvalues $\lambda_k$ are real and negative.

The model defined by Eqs.~(\ref{eq:compactNboxeq}) is meant to describe vertically distributed boxes, but boxes can also be aligned horizontally. This can be useful in order to include the atmosphere over land in the model. In principle we can have interactions between all boxes, making the matrix $\mathbf{K}$ less sparse. The mathematical form of the response function remains the same though, but the characteristic time scales and the weights $b_k$ are changed. Several horizontally distributed boxes could also be useful for modeling a space-dependent depth of the mixed layer.

To make separate boxes for the upper ocean layer and atmosphere over land we adopt the asymmetric heat exchange between land and sea used by \cite{Meinshausen:2011} to obtain the equations 
\begin{align}
C_\text{L} \frac{d\Delta T_\text{L}}{dt} &= -\lambda_\text{L} \Delta T_\text{L} + F_\text{L}(t) + \frac{k}{f_L} (\mu \alpha \Delta T_{1} - \Delta T_\text{L}) \label{eq:landebm}\\
C_{1} \frac{d\Delta T_{1}}{dt} &= -\lambda_\text{O} \Delta T_{1} + F_\text{O}(t) - \frac{k}{f_\text{O}} (\mu \alpha \Delta T_{1} - \Delta T_\text{L}) \label{eq:seaebm} + F_\text{N}. \nonumber
\end{align}
Here it is assumed that the temperature in the atmosphere over oceans $\Delta T_\text{O,atmos}$ is proportional to the temperature in the mixed layer, i.e. $\Delta T_\text{O,atmos} = \alpha  \Delta T_{1}$, where the factor $\alpha>1$ describes the effect of changing sea ice cover \citep{Raper:2001}. The parameter $\mu>1$ quantifies the asymmetry in the heat transport between the atmosphere over the ocean and the atmosphere over land. The parameter $f_\text{L} = 0.29$ is the proportion of Earth's surface that is covered by land, and $f_\text{O} = 1-f_L$. $F_\text{N}$ represents the heat transport into the deep oceans, and $F_\text{L}$ and $F_\text{O}$ are the forcing terms over land and ocean respectively. From the Coupled Model Intercomparison Project  Phase 3 (CMIP3) models one finds that the typical values of $\mu$ are in the range 1 to 1.4 \citep{Meinshausen:2011}. This implies that, when a new equilibrium is reached after a perturbation of the forcing, the land temperature will have changed more than the SST.

In the limit $C_\text{L} \to 0$,  Eq.~(\ref{eq:landebm}) becomes 
\begin{equation}
\Delta T_\text{L} = \frac{F_\text{L}(t) + k \mu \alpha \Delta T_{1} /f_\text{L}}{\lambda_\text{L} + k  / f_\text{L}}. \label{eq:landtemp}
\end{equation}
Hence,  land temperature appears as a weighted sum of the SST and an instantaneous response to the forcing over land. The GMST anomaly is given by
\begin{equation*} 
\Delta T_\text{global} = f_\text{L} \Delta T_\text{L} +f_\text{O} \Delta T_{1}  = \left( f_\text{O} + \frac{k \mu \alpha}{\lambda_\text{L} + k /f_\text{L}} \right) \Delta T_{1} + \frac{f_\text{L}}{\lambda_\text{L} + k /f_\text{L}} F_\text{L}(t). 
\end{equation*}

\section{Approximate scale invariance from aggregation of OU processes} \label{sec:3}

The Ornstein-Uhlenbeck stochastic process is defined via the stochastic differential equation 
\begin{equation*}
dx(t) = -\theta x(t)\,dt + \sigma dB(t), 
\end{equation*}
where $dB(t)$ is the white noise measure. The equation has a stationary solution on the form 
\begin{equation}
x(t) = \int_{-\infty}^t R(t-s)dB(s) \text{~ with ~} R(t) = \sigma e^{-\theta t}. 
\end{equation}
The parameter $\sigma$ is called the scale parameter, and $\theta$ is the damping rate. Since $dB(t)$ is a white noise it follows that 
\begin{equation}
\langle x(t) x(t+\tau) \rangle = \int_0^\infty R(t')R(t'+\tau)dt' = \frac{\sigma^2}{2 \theta} e^{-\theta t}, 
\end{equation}
hence,  the characteristic correlation time of an OU process is $\tau = 1/\theta$.  
In the multi-box EBM with $N$ vertically distributed boxes, the temperature $\Delta T_1(t)$ is given by 
\begin{equation} \label{Tksum}
\Delta T_1(t) = \int_{-\infty}^t \Big{(}\sum_{k=1}^N b_k e^{\lambda_k (t-s)}  \Big{)} d \Delta F(s) = \sum_{k=1}^N b_k \Delta T_{1,k}(t),
\end{equation}
where 
\begin{equation} \label{tk}
\Delta T_{1,k}(t) = \int_{-\infty}^t e^{\lambda_k (t-s)} d\Delta F(s). 
\end{equation}
If we consider the perturbations of the radiative forcing caused by volcanoes, solar variability, and anthropogenic activity as ``deterministic'', and the perturbations from the chaotic atmospheric dynamics as random, then it is natural to model the forcing as a superposition of a deterministic component and a white-noise random process;\footnote{In this paper we follow  \cite{Rypdalx2:2014} and model the random component of the forcing as white noise. However, the models can easily be modified to other stochastic models for the random forcing.}
\begin{equation*}
d\Delta F(t) = \Delta F_\text{det}(t)dt + \sigma dB(t). 
\end{equation*}  
Since the $N$-box models we consider are linear, the decomposition of the forcing yields a straightforward decomposition of the temperature response;
\begin{equation} \label{twoterms}
\Delta T_1(t) = \Delta T_{1,\text{det}}(t) + \sigma \int_{-\infty}^t \Big{(}\sum_{k=1}^N b_k e^{\lambda_k (t-s)}  \Big{)} dB(s) = \Delta T_{1,\text{det}}(t) + \sigma \sum_{k=1}^N b_k x_k(t),
\end{equation}
where the processes  
\begin{equation*}
x_k(t) = \int_{-\infty}^t e^{\lambda_k(t-s)} dB(s)
\end{equation*}
are dependent OU processes with characteristic time scales  given by the eigenvalues of the matrix ${\bf C}^{-1} {\bf K}$ via the relations $\tau_k = -1/\lambda_k$. 
Taking the Fourier transform of Eq.~(\ref{tk}) yields 
\begin{equation*}
  \Delta T_{1,k}(\omega) = \frac{\Delta F(\omega)}{ i \omega +1/\tau_k },
 \end{equation*}
and the PSD of $\Delta T_1(t)$ becomes 
 \begin{equation*}
 S_1(\omega) = \lim_{T \rightarrow \infty} \frac{1}{T} \langle |\Delta T_1(\omega)|^2\rangle =\lim_{T \rightarrow \infty} \frac{1}{T} \langle  |\Delta F(\omega)|^2\rangle  \left[ S^{(0)}(\omega) + S^{(\text{cr})}(\omega) \right],
\end{equation*}
 where 
 \begin{equation} \label{spect}
 S^{(0)}(\omega) =  \sum_k  \frac{b_k^2}{ \left( \omega_k^2+ \omega^2 \right)} \text{~,~ }
 S^{(\text{cr})}(\omega) =
\sum_k \sum_{ j<k}\frac{2b_kb_j  \left( \omega_k\omega_j + \omega^2\right)}{\left(\omega_j^2+\omega^2\right)\left(\omega_k^2+\omega^2\right)  },
\text{~ and~ }  \omega_k=1/ \tau_k. 
\end{equation}
Here $T$ is the length of the time series $\Delta T_1(t)$, $ S^{(0)}(\omega)$ is the PSD of an independent superposition of the processes $\Delta T_{1,k}(t)$, and $S^{(\text{cr})}(\omega)$ is the contribution to the PSD from the cross terms, which can not be neglected since the processes $\Delta T_{1,k}(t)$ are driven by the same forcing $\Delta F(t)$. For the stochastic component of the process Eq.~(\ref{twoterms}) we can replace the forcing by a white noise process,  such that $\lim_{T \rightarrow \infty} \frac{1}{T} \langle  |\Delta F(\omega)|^2\rangle$ is a constant in $\omega$. 

\begin{figure}
\includegraphics[width=12cm]{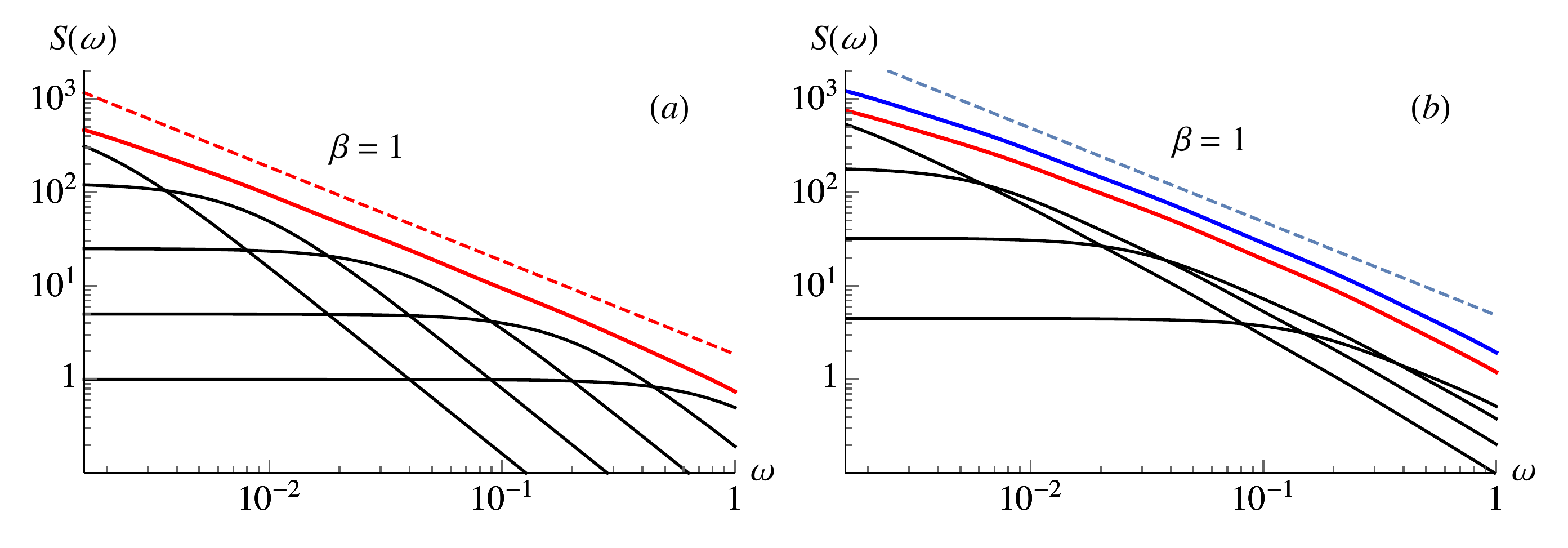}
\caption{The solid red line in (a) shows the sum $S^{(0)}(\omega)$ in Eq.~(\ref{spect}), with $N=5$, $\beta=1$, $\tau_1 = 1$, $b_1=1$ and $a=5$. The dashed red line shows a power law $1/f^\beta$ with $\beta=1$. The black lines show the contribution from each term in the aggregation. The red line in (b) shows the sum $S^{(\text{cr})}(\omega)$, while the blue line shows $S^{(0)}(\omega) + S^{(\text{cr})}(\omega)$. The dashed blue line shows the power law spectrum we approximate, given the same forcing for all modes. The black lines in (b) show the contribution to $S^{(\text{cr})}(\omega)$ for each $k$: $\sum_{ j<k}\frac{2b_kb_j  \left( \omega_k\omega_j + \omega^2\right)}{\left(\omega_j^2+\omega^2\right)\left(\omega_k^2+\omega^2\right)}$.}
\label{fig:5modes}
\end{figure}

The PSD in Eq.~(\ref{spect}) can easily be made to approximate a power law. For instance we can pick time scales $\tau_k$ such that $\tau_{k+1} = a \tau_k$ and weights $b_k$ such that $b_{k+1} = \sqrt{a^{\beta - 2}} b_k$. Then we have the approximate relations $S^{(0)}(\omega) \sim \omega^{\beta}$, $S^{(\text{cr})}(\omega) \sim \omega^{\beta}$ and hence  $S_1(\omega) \sim \omega^{\beta}$. In Fig.~\ref{fig:5modes} this is demonstrated for a superposition of $N=5$ terms. The sum $S^{(0)}(\omega)$ is shown by the red line in (a) and the total sum $S^{(0)}(\omega) + S^{(\text{cr})}(\omega)$ by the blue line in (b). The idea that a long memory process can be produced by aggregating OU processes is the same as presented by e.g. \cite{Granger:1980} and \cite{MRypdal:2016}. In this paper the time scales $\tau_k$ and their weights $b_k$ are estimated from data without {\em a priori} assumptions of scale invariance.

\subsection{Example 1. The two-box model}

\begin{figure}
\begin{center}
\includegraphics[width=15cm]{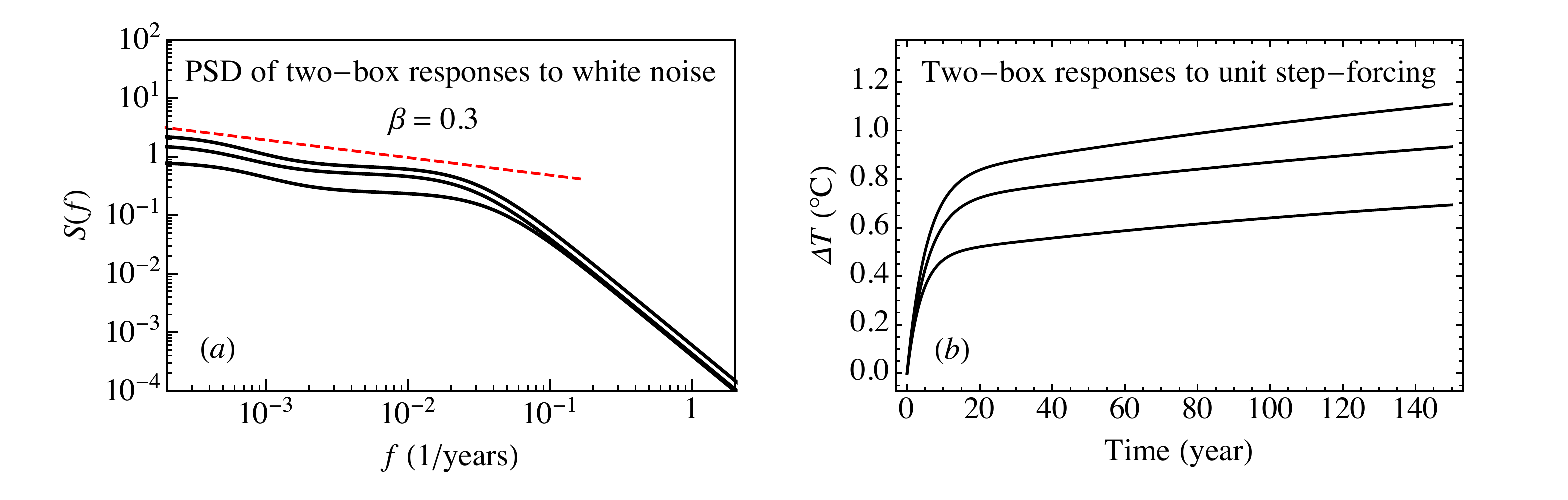}
\end{center}
\caption{(a) Theoretical PSD of the response to white-noise forcing for a two-box model, with three different sets of parameters. (b) The response to a unit-step forcing for the same parameter choices. The three different sets of parameter choices are those estimated by \cite{Geoffroy:2013one} from the following models: HadGEM2-ES:  $(\kappa_1,\kappa_2) = (0.65, 0.55) \text{ W m}^{-2} \text{K}^{-1}$, $(C_1, C_2) = (6.5,82) \text{ W yr m}^{-2} \text{K}^{-1}$, IPSL-CM5A-LR: $(\kappa_1,\kappa_2) = (0.79, 0.59) \text{ W m}^{-2} \text{K}^{-1}$, $(C_1, C_2) =(7.7, 95) \text{ W yr m}^{-2} \text{K}^{-1}$, NorESM1-M: $(\kappa_1, \kappa_2) = (1.11, 0.88) \text{ W m}^{-2} \text{K}^{-1}$, $(C_1, C_2) = (8.0,105)\text{ W yr m}^{-2} \text{K}^{-1}$. 
}
\label{fig:twoboxmodels}
\end{figure}

For the classical two-box model, \cite{Geoffroy:2013one} estimated parameters by fitting the GMST-response in the two-box model to the corresponding response in CMIP5 models. The forcing scenario used for fitting were abrupt quadrupling of atmospheric CO$_2$ concentration.
They find multi-model mean parameter estimates $\hat{C}_1 = 7.3$ W yr m$^{-2}$  K$^{-1}$, $\hat{C}_2 = 106$ W yr m$^{-2}$  K$^{-1}$, $\hat{\kappa}_1 = 1.13$ Wm$^{-2}$K$^{-1}$, $\hat{\kappa}_2 = 0.73$ Wm$^{-2}$K$^{-1}$, which correspond to characteristic time scales of $3.88$ years and $242$ years. The two-box model provides a good fit to CMIP5 abrupt 4$\times$CO$_2$ experiments and 1$\%$ per year CO$_2$ increase experiments over 140 years, but if forced with white noise, the PSD is $S_1(\omega) \propto S^{(0)}(\omega) + S^{(\text{cr})}(\omega)$. Using the parameters estimated by \cite{Geoffroy:2013one} for three different CMIP5 models, we have plotted this expression in Fig.~\ref{fig:twoboxmodels}(a). As seen from the figure, the PSD can be approximated by two different power laws; one  in the high-frequency range and another in the low-frequency range. For frequencies corresponding to time scales greater than a few decades the PSD can be approximated by $S(f) \propto 1/f^\beta$ with $\beta=0.3$, and for the higher frequencies it can be approximated by $S(f) \propto 1/f^2$.  
This result is inconsistent with the PSDs estimated from CMIP5 control run temperatures, which are well approximated by one power-law over the entire range of frequencies from months to centuries \citep{Fredriksen:2016}. 
 
The inability of the two-box model to simultaneously describe the the average GMST response to certain forcing scenarios in GCMs, as well as the PSD of the background fluctuations, serves as a motivation to analyze more general $N$-box models, and in the next example we consider the EBM with three vertically distributed boxes. We will demonstrate that the three-box model provides accurate descriptions of both the ``deterministic'' response to historic radiative forcing and the statistical properties of the response to random forcing.

\subsection{Example 2. The three-box model}

The three-box model is given by Eq.~(\ref{eq:compactNboxeq}) with 
\begin{equation*}
\mathbf{C} = 
 \begin{pmatrix}
  C_1 & 0 & 0 \\
  0 & C_2 & 0 \\
  0 & 0 & C_3
 \end{pmatrix}, \text{~}
\Delta \mathbf{T} =  \begin{pmatrix}
\Delta T_1 \\
\Delta T_2 \\
\Delta T_3
 \end{pmatrix}, \text{~}
 \Delta \mathbf{F(t)} =\begin{pmatrix}
\Delta F(t) \\
0 \\ 0
\end{pmatrix}
 \end{equation*}
 and 
\begin{equation} \label{K}
 \mathbf{K} =  \begin{pmatrix}
 -(\kappa_1 + \kappa_2) & \kappa_2  & 0 \\
  \kappa_2 & -(\kappa_2 + \kappa_3) & \kappa_3 \\
  0 & \kappa_3 & -\kappa_3
 \end{pmatrix}. 
\end{equation}

To estimate the parameters in the model we will make use of the HadCRUT4 data set for the GMST since 1850 \citep{Morice:2012}, and the global effective forcing data, both with annual resolution. The forcing data is an updated version of \cite{Hansen:2011}, available at \url{http://www.columbia.edu/~mhs119/Forcings/}.
We also use the Moberg Northern Hemisphere reconstruction \citep{Moberg:2005}, and the Crowley forcing data \citep{Crowley:2000} for the years 1000-1979. We fix a set of three well separated time scales ($\tau_1$, $\tau_2$ and $\tau_3$) and compute the responses 
\begin{equation*} 
\Delta T_{1,k}(t) = \int_{t_0}^t e^{-(t-s)/\tau_k} \Delta F(s) ds \text{~ for ~} k=1,2 \text{ and } 3,  
\end{equation*}
to the historical forcing data $\Delta F(t)$, where the integral is estimated by a sum. As in Eq.~(\ref{Tksum}), the GMST response is a linear combination of the responses $\Delta T_{1,k}(t)$: 
\begin{equation*}
\Delta T_{1}(t) = b_1 \Delta T_{1,1}(t) + b_2 \Delta T_{1,2}(t) + b_3 \Delta T_{1,3}(t), 
\end{equation*}
and our approach is to estimate the parameters $b_1$, $b_2$ and $b_3$ from historical data of GMST and forcing. We will subsequently demonstrate that for the range of time scales we consider in this paper, the results are largely insensitive to the choice of time scales ($\tau_1$, $\tau_2$ and $\tau_3$), as long as these are sufficiently separated. We do not only require that the ``deterministic'' response to radiative forcing fits well with observations, but also that the PSD of the stochastic component of the response is consistent with the estimated PSD of the residual observational signal (the difference between observed temperature record and the model response to the ``deterministic'' forcing). \cite{Rypdalx2:2014} employ a maximum-likelihood method to estimate the variance of the temperature residual $\sigma_T^2$ and the spectral exponent $\beta$ of a long-memory model. This method is inadequate here, since we have more free parameters and the method favorizes a good fit for the smallest time scales. 

We employ instead an iterative routine, which in addition to weighting all the time scales in the estimation, allows us to fit to a composite spectrum. We first compute the residual temperature by guessing the model parameters, computing the deterministic responses to the Crowley and Hansen forcing time series, and then subtracting the deterministic responses from the observed Moberg and HadCRUT4 temperature records. The PSD of the residual is estimated using the periodogram, and subsequently log-binned. The theoretical PSD of the response to white-noise forcing in three-box model is then fitted to the composite residual PSD derived  from the Moberg and  HadCRUT4 temperatures by minimizing the square distance between the theoretical and estimated residual PSDs. From this we obtain new estimates of the relative size of the weights $b_k$ at each time scale, and using these we perform a new regression analysis with HadCRUT4 to determine the initial temperature $T_0$ and the absolute strength of the response. The procedure is repeated with these new model parameters as a starting point. The parameter estimation converges rapidly; only a few iterations are needed to obtain our estimates. 
 
\begin{figure}[t]
\includegraphics[width=13cm]{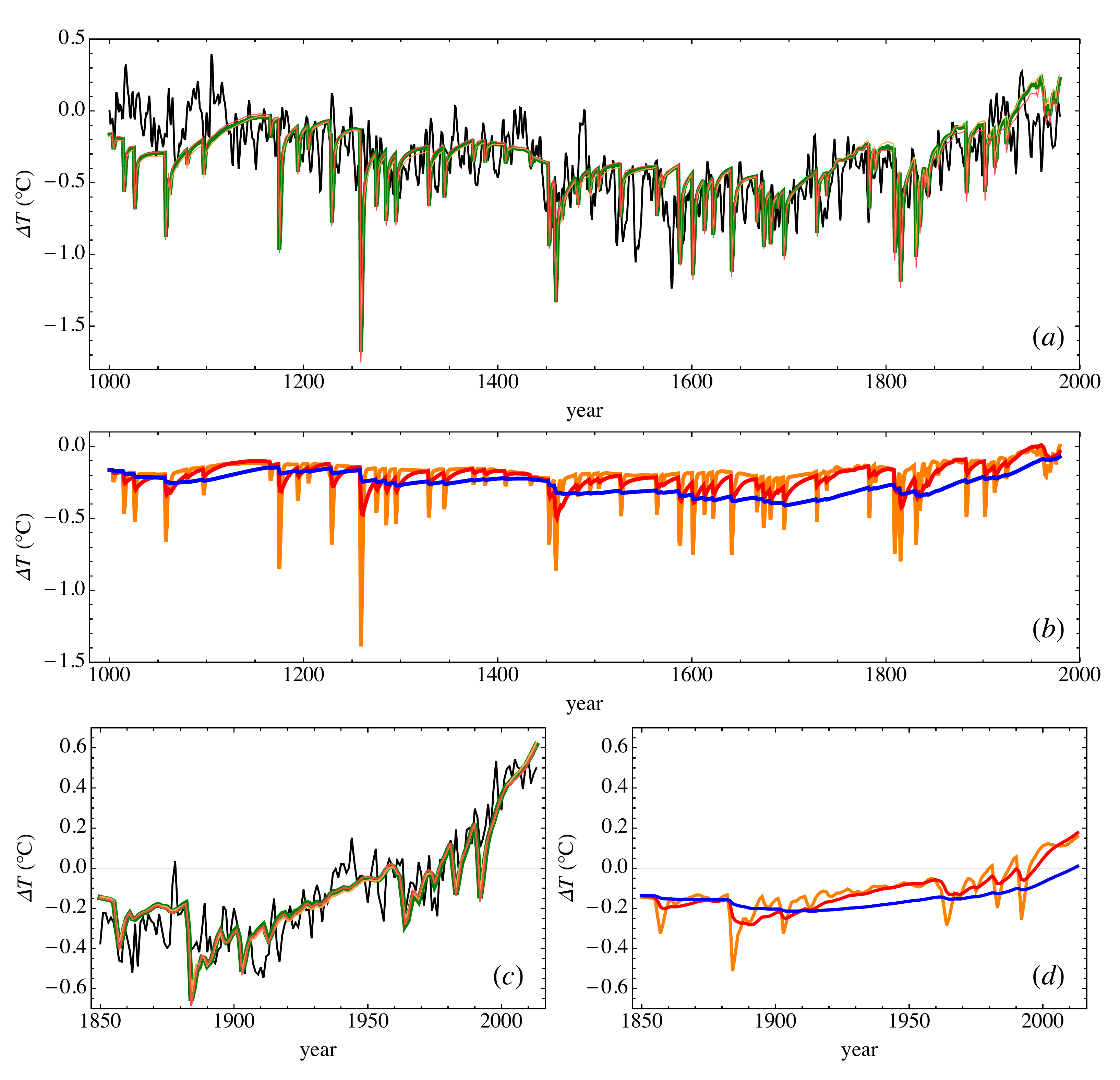}
\caption{(a) The black curve is the Moberg reconstruction, and the colored curves are the responses to Crowley forcing of the three-box model with estimated parameters. The orange curve is the response of the three-box model where we have fixed time scales $\tau_1 = 0.5$ yrs, $\tau_2 = 5$ yrs  and $\tau_3 = 50$ yrs. The thicker green curve is the response of the three-box model where we have fixed time scales $\tau_1 = 1$ yr, $\tau_2 = 10$ yrs  and $\tau_3 = 100$ yrs. The red curve is the response of the three-box model where we have fixed time scales $\tau_1 = 1$ yr, $\tau_2 = 20$ yrs  and $\tau_3 = 400$ yrs. (c) is similar to (a), but shows response to forcing over the instrumental period, and the black curve is the global instrumental temperature HadCRUT4. (b) and (d) show the green curves in (a) and (c) decomposed into the responses corresponding to the exponential terms in the response function $R(t)$.  The orange curve corresponds to $\tau_1=1$ yr, the red curve corresponds to $\tau_2 =10$ yrs, and the blue curve corresponds to $\tau_3$ = 100 yrs.}
\label{fig:responses}
\end{figure}

In Fig.~\ref{fig:responses}(a) and Fig.~\ref{fig:responses}(c) we show the deterministic responses to historical forcing with estimated parameters. The three different colors represent different choices of the time scales $\tau_1$, $\tau_2$ and $\tau_3$, and the parameter estimates are presented in Table~\ref{tab:parameters}, where we also present the corresponding values of the parameters $\kappa_k$ and $C_k$, as well as the equilibrium climate sensitivity of the model. We note that the colored curves in Fig.~\ref{fig:responses}(a) and Fig.~\ref{fig:responses}(c) are almost indistinguishable, and they all closely follow the HadCRUT4 and Moberg records. 

\begin{figure}
\includegraphics[width=14cm]{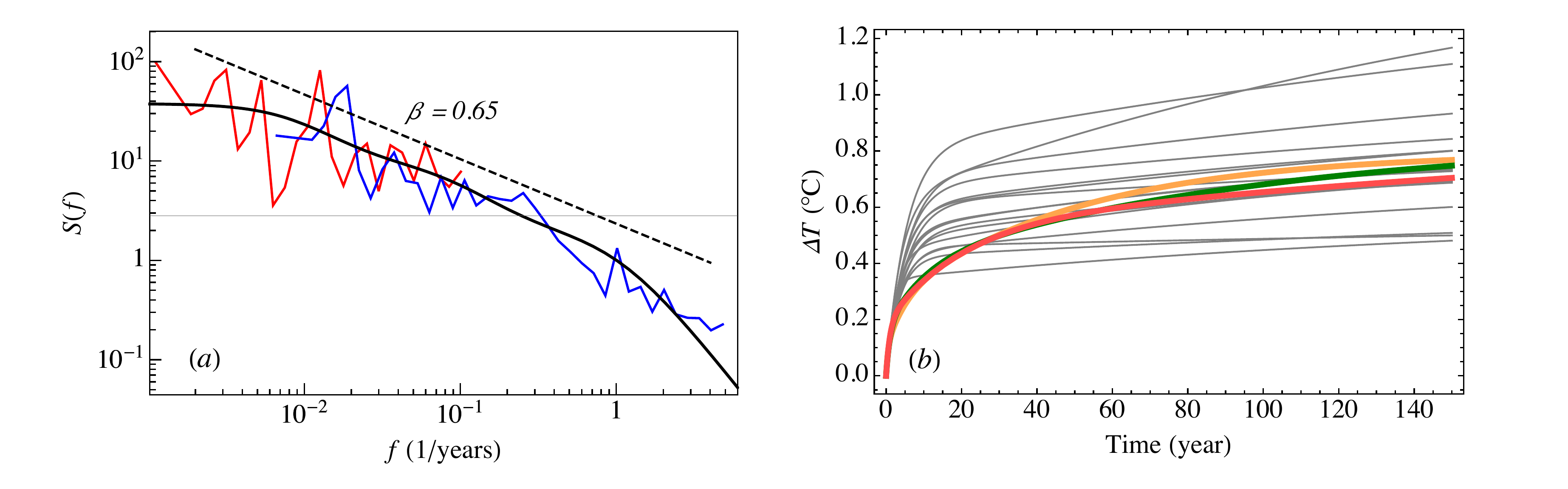}
\caption{(a) The blue curve is the estimated PSD of the residual global instrumental temperature after subtracting the estimated deterministic response of a three-box model. The characteristic time scales in the three-box model are chosen to be $\tau_1=1$ yr, $\tau_2=10$ yrs and $\tau_3=100$ yrs. The red curve is the residual of the Moberg reconstruction. Both curves are normalized by their power on decadal scales. The black curve is the theoretical PSD with the estimated response function, which can be quite well approximated by a power-law, shown by the dashed line with slope $-\beta = -0.65$. (b) The gray curves are the responses to a unit step-forcing using two-box model parameters estimated for many climate models by \cite{Geoffroy:2013one}. The colored curves show the three-box responses with the parameters from Table \ref{tab:parameters}. The orange curve is the response of the three-box model where we have fixed time scales $\tau_1 = 0.5$ yrs, $\tau_2 = 5$ yrs  and $\tau_3 = 50$ yrs. The thicker green curve is the response of the three-box model where we have fixed time scales $\tau_1 = 1$ yr, $\tau_2 = 10$ yrs  and $\tau_3 = 100$ yrs. The red curve is the response of the three-box model where we have fixed time scales $\tau_1 = 1$ yr, $\tau_2 = 20$ yrs  and $\tau_3 = 400$ yrs.}
\label{fig:resspectra_stepresponses}
\end{figure}

Fig.~\ref{fig:resspectra_stepresponses}(a) shows the theoretical PSD of the stochastic component (the response to white noise forcing) in the model. The estimated parameters are shown in Table \ref{tab:parameters}, and the choice of time scales are $\tau_1=1$ yr, $\tau_2=10$ yr and $\tau_3=100$ yr. The PSD of the model fits well with the PSD estimated from observational data in the time-scale range from months to centuries, and in this range it is close to a power-law with an exponent $\beta = 0.65$.

In Fig.~\ref{fig:resspectra_stepresponses}(b) we show the three-box model responses to a unit step-forcing scenario. The different colors correspond to different choices of the time scales $\tau_1$, $\tau_2$ and $\tau_3$, and the gray curves are the corresponding two-box model responses with parameters estimated by \cite{Geoffroy:2013one} by fitting to abrupt 4$\times$CO$_2$ experiments in CMIP5 models. From the response to a unit-step forcing we can also derive that the equilibrium climate sensitivity for an $N$-box model is given by $S_\text{eq} = \sum_{k=1}^N b_k \tau_k$.

The difference we observe in the step-forcing responses is not a result of a difference between the two- and three-box models, but rather a difference in estimation strategy. In our estimation procedure, we use only observational and proxy data, and have hence chosen the parameters that best reproduce both the residual spectra and responses to historical forcing. The parameters estimated by this method are not the same as the parameters that best describe the 4$\times$CO$_2$ runs. This discrepancy can have several explanations; perhaps is the random forcing not accurately modeled as a {\em white} noise, or the forcing in the 4$\times$CO$_2$ experiments is too strong for a linear approximation to be valid. It is likely that the feedback parameter or parameters related to ocean mixing can change during the strong and abrupt climate change following a quadrupling in CO$_2$ concentration. The large differences between the different CMIP5 step-responses also reflect the large uncertainty associated with these model runs.

\color{black}

\begin{table*}[t]
\caption{Parameters estimated from data. The time scales $\tau_k$ and their weights $b_k, \;k=1,2,3 $ uniquely determine the estimated response functions. The parameter $\sigma_{\text{monthly}}$ is the estimated standard deviation of the monthly resolved stochastic forcing. $T_{0,\text{instr}}$ and $T_{0,\text{moberg}}$ are the initial temperature anomalies estimated for global instrumental temperature and the Moberg reconstruction. We note that there is some remnant seasonal variability that we were not able to remove by subtracting a mean seasonal cycle. This is not so apparent in a double-logarithmic plot of the PSD, but it causes our estimates of $\sigma_{\text{monthly}}$ to be slightly inconsistent with the actual residual variability. We also point out that our estimates of $\sigma_{\text{monthly}}$ are only valid for monthly resolved temperatures, or monthly temperatures sampled at a different time resolution. }
\begin{center}
\begin{tabular}{lcccc}
\hline \hline
 & Unit & & & \\ 
$(\tau_1, \tau_2, \tau_3)$ & yr & (0.5, 5, 50) & (1, 10, 100) & (1, 20, 400) \\ \hline
$b_1$ & $\text{K m}^2 \text{ W}^{-1} \text{ yr}^{-1}$ & 0.198 & 0.165 & 0.187 \\
$b_2$ & $\text{K m}^2 \text{ W}^{-1} \text{ yr}^{-1}$ & 0.033 & 0.022 & 0.018 \\
$b_3$ & $\text{K m}^2 \text{ W}^{-1} \text{ yr}^{-1}$ & 0.011 & 0.005 & 0.001 \\
$\sigma_{\text{monthly}} $ & $\text{W m}^{-2} \text{yr}^{-1/2} $ & 0.73 & 0.70 & 0.68 \\
$T_{0,\text{instr}}$ & K & -0.13 & -0.14& -0.14 \\
$T_{0,\text{moberg}}$ & K & -0.18 & -0.17 & -0.15 \\
$S_\text{eq}$ & $ \text{K m}^2 \text{ W}^{-1}$ & 0.79 & 0.85 & 1.07 \\ \hline
$C_1$ & $\text{W yr m}^{-2} \text{ K}^{-1}$ & 4.15 & 5.09 & 4.86 \\
$C_2$ & $\text{W yr m}^{-2} \text{ K}^{-1}$ & 15.4 & 22.0 & 31.7 \\
$C_3$ & $\text{W yr m}^{-2} \text{ K}^{-1}$ & 24.1 & 41.8 & 149 \\
$\kappa_1$ & $\text{W m}^{-2} \text{ K}^{-1}$ & 1.26 & 1.18 & 0.94 \\
$\kappa_2$ & $\text{W m}^{-2} \text{ K}^{-1}$ & 5.61 & 3.30 & 3.50 \\
$\kappa_3$ & $\text{W m}^{-2} \text{ K}^{-1}$ &1.74 & 1.2 & 0.91 \\
\hline
\end{tabular}
\end{center}
\label{tab:parameters}
\end{table*}

\begin{figure}
\includegraphics[width=12cm]{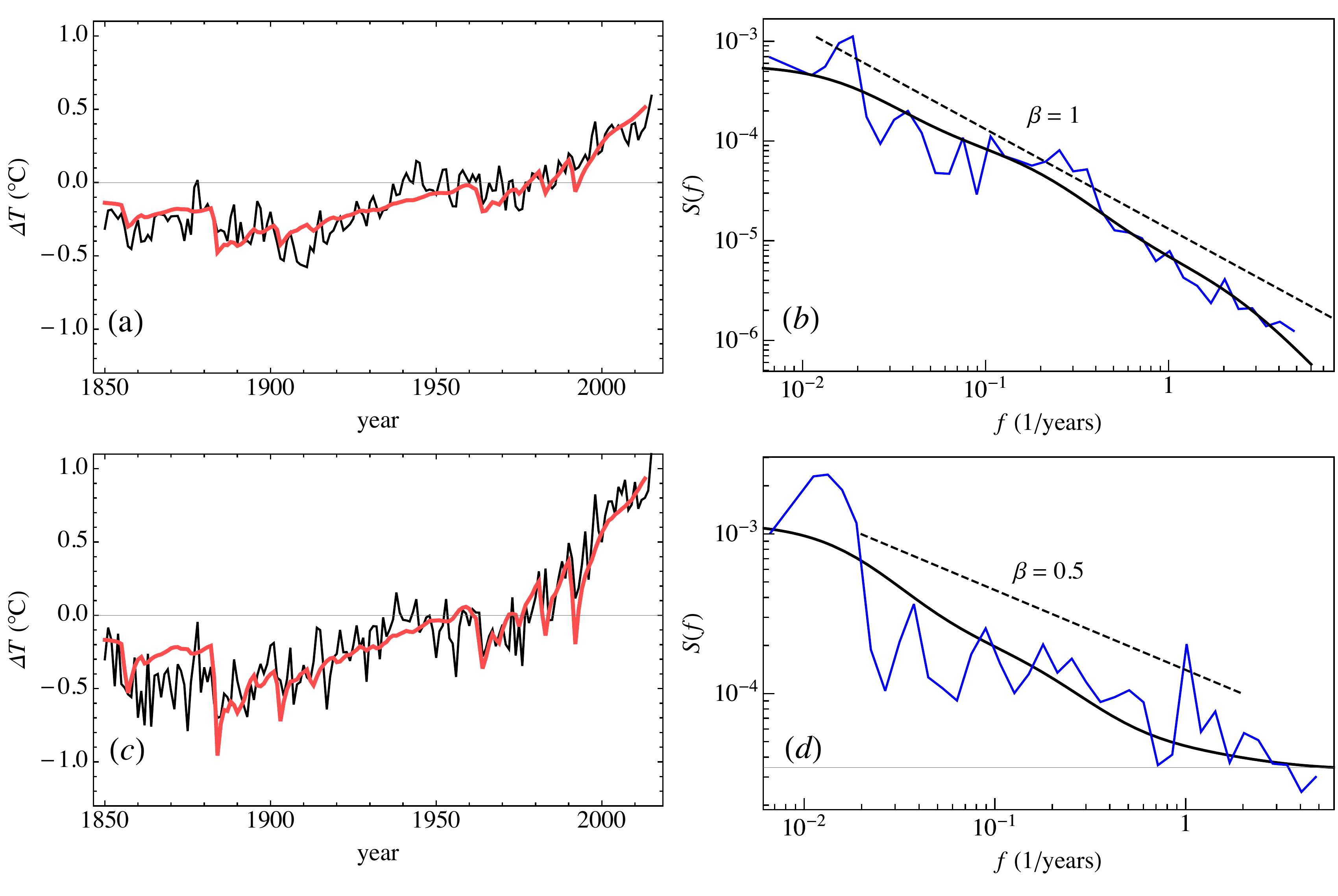}
\caption{(a) The black curve is the global SST dataset HadSST3, and the red curve the response to forcing on time scales of 0.5, 5 and 50 years. (b) The blue curve is the spectrum of the residual of HadSST3 after subtracting the red in (a), and the black curve is the expected spectrum of the residual. (c) and (d) Similar to (a) and (b), but for the global land temperature dataset CRUTEM4v. In addition to the three time scales, we have an instantaneous response to the forcing for land temperatures.}
\label{fig:landandsearesponses}
\end{figure}

\begin{table}[t]
\caption{
Parameters estimated from global sea and land surface temperatures. The fixed time scales are chosen to be $\tau_1 = 0.5$ yrs, $\tau_2=5$ yrs and $\tau_3=50$ yrs.}
\begin{center}
\begin{tabular}{llllll}
\hline \hline
sea:  & & & land: & & \\ \hline
$b_1$ & 0.092 & $\text{K m}^2 \text{ W}^{-1} \text{ yr}^{-1}$ & $r_1$ & 0.10 & $\text{K m}^2 \text{W}^{-1}$\\
$b_2$ & 0.035 & $\text{K m}^2 \text{ W}^{-1} \text{ yr}^{-1}$ & $r_2$ & 1.40 & \\
$b_3$ & 0.009 & $\text{K m}^2 \text{ W}^{-1} \text{ yr}^{-1}$ & & & \\
$\sigma_{\text{monthly,sea}} $ & 0.76 & $\text{W m}^{-2} \text{yr}^{-1/2}$ & $\sigma_{\text{monthly,land}}$ & 0.2 & $\text{W m}^{-2} \text{yr}^{-1/2} $ \\
$T_{0,\text{sea}}$ & -0.13 & K & $T_{0,\text{land}}$ & -0.15 & K \\
$S_{\text{eq,sea}}$ & 0.69 & $ \text{K m}^2 \text{ W}^{-1}$ & $S_{\text{eq,land}}$ &  1.07 & $ \text{K m}^2 \text{ W}^{-1}$\\
\hline
\end{tabular}
\end{center}
\label{tab:landandseaparameters}
\end{table}

\subsection{Example 3. Separate boxes for land and ocean}

With two surface boxes, one for land and one for ocean, the equation for the ocean surface temperature is 
\begin{equation}
C_\text{1} \frac{d}{dt} \Delta T_\text{1} = - \kappa_1' \Delta T_\text{1} + F_\text{N} + F_\text{O}(t) + \frac{k}{f_\text{O}\left( \lambda_\text{L} + k/f_\text{L} \right) }F_\text{L}(t) 
\label{eq:upperoceaneq}
\end{equation}
where 
\begin{equation}
\kappa_1' = \lambda_\text{O} + k \mu \alpha /f_\text{O} - \frac{k^2 \mu \alpha}{f_\text{O} f_\text{L} \left(\lambda_\text{L} + k /f_\text{L} \right)}. 
\end{equation}
The response function can hence be estimated in the same way as for global temperature, and the results are given in Table \ref{tab:landandseaparameters}. For the estimation we have used the global SST dataset HadSST3 \citep{Kennedy:2011one, Kennedy:2011two}.

Simplifying the constants in Eq. (\ref{eq:landtemp}), and separating the stochastic and deterministic parts results in
\begin{equation}
\Delta T_\text{L}(t) = r_1 F_\text{L}(t) + r_2 \Delta T_{1} = r_1 F_\text{L,det}(t) + r_2 \Delta T_\text{1,det}(t) + r_1 \sigma_\text{L} dB_\text{L}(t) + r_2 \Delta T_\text{1,stoc}(t),
	\label{eq:landcomps}
\end{equation}
where $\sigma_\text{L} dB_\text{L}(t)$ is the direct stochastic forcing of the land surface temperature. For the ocean response we use the previously estimated parameters, given in Table \ref{tab:landandseaparameters}. The remaining parameters are chosen such that the deterministic response
\begin{equation}
\Delta T_\text{L,det} = T_0 + r_1 F_\text{L,det} + r_2 \Delta T_\text{1,det}
\label{eq:detlresponse}
\end{equation}
is similar to global land surface temperature (LST), and such that the PSD of the residual temperature obtained after subtracting the deterministic response is similar to the PSD expected for the stochastic part of Eq. (\ref{eq:landcomps}):
\begin{equation}
 \langle |\Delta T_\text{L,stoc}(\omega) |^2 \rangle = (r_1 \sigma_\text{L})^2 \Delta t + r_2^2  \langle |\Delta T_\text{1,stoc}(\omega) |^2 \rangle
 \label{eq:stocspectrum}
\end{equation}
Here we assume that the instantaneous response in ocean temperature to changes in the direct forcing of the land temperature is small compared to the land temperature response to this forcing. For the global LST we use the CRUTEM4v dataset \citep{Jones:2012}. The observed SST, LST and the response to deterministic forcing with the parameters listed in Table \ref{tab:landandseaparameters} are shown in Fig.~\ref{fig:landandsearesponses}. This figure also shows the estimated and theoretical PSD of the response to stochastic forcing with the same parameters. The global temperature response $\Delta T_\text{G} = f_\text{L} \Delta T_\text{L} +f_\text{O} \Delta T_{1} $ is similar to the three-box temperature response estimated directly from global temperature.

The theoretical PSD of $\Delta T_1(t)$ fits well with the estimated PSD of the SST residual, and both are well-approximated by a power law with $\beta_\text{O} \approx 1$. In the same way, the theoretical PSD of $\Delta T_\text{L}(t)$ fits well with the estimated PSD of the LST, and both can be approximated by a power law with $\beta_\text{L} \approx 0.5$.
These results are similar to the estimated PSDs of the linearly de-trended global LST and global SST analyzed in \cite{Fredriksen:2016}. With this model, the only reason global LST shows persistence is because of the influence by global SST, but the persistence is weaker for land than for sea due to the component responding instantly to forcing.

We note that in the model presented in this paper, the relation between the equilibrium climate sensitivities for land temperatures and ocean temperatures is
\begin{equation*}
S_\text{eq,land} = r_1 + r_2 S_\text{eq,sea}.
\end{equation*} 
If $r_1/r_2 \ll S_\text{eq,sea}$, there will be a near constant ratio between land and ocean temperature change, consistent with the findings of \cite{Lambert:2011}.

\section{Concluding remarks} \label{sec:4}
Simple climate models can be divided in two classes: EBMs and tuned impulse-response (IR) linear statistical models \citep{Good:2011}. The power-law response model proposed by \cite{Rypdalx2:2014} is an example of an IR model that reproduces observed temperature variability quite well, but in this mathematical idealization conservation of energy is lost. Some may also find it problematic that it is not derived directly from physics. In this paper we demonstrate that such a model is closely approximated by the response of a multi-box EBM. This shows that LRD models can be seen as a compact mathematical descriptions of the effect of simple and concrete physical mechanisms. 

Using only three boxes, we can produce a spectrum of the internal temperature variability that is practically indistinguishable from a power-law spectrum over a frequency range spanning more than two orders of magnitude. The response closely approximates a power-law response, even if we do not assume this \textit{a priori}. The results of this paper are consistent with the findings of \cite{Fraedrich:2004} and \cite{Lemke:1977}, who demonstrate $1/f$-noise characteristics in linear diffusion models.  

As suggested by \citep{Ragone:2015}, we belive that the global temperature response in the Holocene can be well approximated as linear. Extending the linear response to local and regional temperatures is more problematic, especially for the temperature responses in the Arctic, where strong non-linear effects such as the sea-ice albedo feedback are present.
However, the multi-box models can easily be extended to include non-linear terms, for instance to describe the rapid sea ice loss in the Arctic. They can also be extended to include different types of tipping points, and this allows ut to study critical transitions in systems that exhibit LRD. The effect of LRD on the so-called early-warning indicators associated with critical transitions in multi-box EMBs is a topic that will be pursued in future work.

\acknowledgments
The authors thank K. Rypdal and O. L{\o}vsletten for useful discussions. 
The paper was supported by the Norwegian Research Council (KLIMAFORSK programme) under grant no. 229754.

\end{document}